\documentclass[prb,twocolumn,showpacs,amsmath,amssymb,superscriptaddress]{revtex4}

\usepackage{graphicx}
\usepackage{dcolumn}
\usepackage{bm}
\usepackage{subfigure}

\newcommand{\be}{\begin{equation}}
\newcommand{\ee}{\end{equation}}
\newcommand{\bea}{\begin{eqnarray}}
\newcommand{\eea}{\end{eqnarray}}
\newcommand{\mbf}{\mathbf}

\begin{document}

\title{Effects of disorder on the vortex charge}
\author{J. Lages}
\email{jose.lages@univ-fcomte.fr} \affiliation{Laboratoire de
Physique Mol\'eculaire, UMR 6624 du CNRS, Universit\'e de
Franche-Comt\'e, La Bouloie, 25030 Besan\c con Cedex, France}
\affiliation{Laboratoire de Physique Th\'eorique, UMR 5152 du CNRS,
Universit\'e Paul Sabatier, 31062 Toulouse Cedex 4, France }
\author{P. D. Sacramento}
\email{pdss@cfif.ist.utl.pt}
\affiliation{
Centro de F\'\i sica das Interac\c c\~oes Fundamentais,
Instituto Superior T\'ecnico,
Av. Rovisco Pais, 1049-001 Lisboa, Portugal}

\date{\today}

\begin{abstract}
We study the influence of disorder on the vortex charge, both due to random pinning
of the vortices and due to scattering off non-magnetic impurities.
In the case when there are no impurities
present, but the vortices are randomly distributed, the effect is
very small, except when two or more vortices are close by.
When impurities are present,
they have a noticeable effect on the vortex charge. This, together
with the effect of temperature, changes appreciably the vortex
charge. In the case of an attractive impurity potential the sign
of the charge naturally changes.
\end{abstract}


\maketitle

\section{Introduction}

Some time ago \cite{Khomskii} it was proposed that the vortex induced
by an external magnetic field in a type-II superconductor should be
electrically charged. This effect was proposed to occur since the
chemical potential is expected to be larger in the vortex core than in
the bulk of the superconductor. It is energetically favorable for
the electrons to lower their energy through the condensation energy
and, since the vortex core is interpreted as being a normal region,
the electrons tend to move to the bulk leaving a charge deficiency
close to the vortex line.
It is the electrochemical potential (the sum
of the chemical potential and electrostatic energy) that is constant
in the superconductor. Due to the circulating currents around the vortex
line the electrostatic potential is needed to compensate the centrifugal
force due to the circular motion \cite{London}.

We should note that the initial understanding that the vortex core region is
populated by normal electrons
has been questioned.
In particular, in the case of clean superconductors, where the mean free path is much larger
than the coherence length, the localized states bound to the vortex core are the result
of Andreev scattering \cite{96}. The core states are coherent superpositions of particle
and hole states and interpreted as being the result of constructive interference
of multiple Andreev scattering
from the spatial variation of the order parameter.
Also it was shown that the main contribution to the supercurrent is originated in these
states.

Soon after the proposal of Khomskii and Freimuth it was suggested that the effect
could be tested experimentally due to the dipole field created at the
surface of the superconductor \cite{Blatter}.
It has been claimed that the charge of the vortex has been measured
using NMR in high temperature superconductors \cite{Kumagai}.

Theoretical studies on the existence of the vortex charge were carried
out subsequently \cite{Hayashi,Machida1,Goryo,Kolacek}. In particular, a relation
was established between the vortex core charge and the vortex bound states
for an s-wave vortex \cite{Hayashi}.
In particular, in the quantum limit
the influence of the bound states is important.
This regime is reached when $T/T_c\leq 1/(k_F \zeta_0)$, where
$T_c$ is the critical temperature, $k_F$ is the Fermi momentum and $\zeta_0=v_F/\Delta_0$
is the coherence length, corresponding to a regime where the thermal width is smaller
than the level spacing. In this
regime the particle-hole asymmetry in the local density of states (LDOS) has been related
to the vortex charge \cite{Hayashi}. The asymmetry results from the different
effect the supercurrent around the vortex has on the particle and hole
wave functions \cite{Berthod}.
However, in d-wave superconductors
the low lying states are not localized \cite{FT1}.
As shown recently, it is the winding of the phase around the vortex and not the detailed
decreasing of the gap amplitude near the vortex core that is ultimately responsible
for the supercurrent and the nature of the low-lying states (bound states in the
$s$-wave case) and their effect on many properties \cite{Berthod,Cardoso}, such as the vortex
charge.

Taking into account the screening
of the vortex charge, the Friedel oscillation in the charge profile was
obtained for the case of a s-wave vortex \cite{Machida1}, showing that
the charge is screened but prevails with a somewhat reduced value.
Other pairing symmetries were also considered \cite{Goryo,Chen} showing
that the existence of the vortex charge is universal.
In all these studies the vortex is charged positively (electron defficiency).
This positive charge has been argued to be the cause of the Hall anomaly
where the Hall conductance changes sign when entering the superconducting
phase \cite{Dorsey}.

In the high magnetic
field regime, where a Landau level description is appropriate, the low
lying states are coherent through the vortex lattice \cite{Dukan} both
in the s-wave \cite{Dukan} and in the d-wave cases \cite{Yasui}.
The coherent nature of the states originates gapless superconductivity
due to the center of mass motion of the Cooper pairs and the action of
a high magnetic field \cite{Dukan,PRL1}. In this regime the vortex charge
has not yet been studied.

In d-wave superconductors
it is likely that other orderings compete with the superconductivity. In
particular, antiferromagnetism \cite{SO5}, or d-density waves \cite{Laughlin}.
The vortex structure and, in particular, the vortex charge, have also been studied
when there is competition between the various order parameters
\cite{Franz,Chen,Chen2,Maska,Kallin}.
In general, a small
region around the vortex will have a non-vanishing order parameter which
affects the density of states and in particular the vortex charge.
It may change sign from an electron defficiency to an electron
abundance at the vortex core. Also, increasing the temperature, in a regime
of parameters where the competing order is absent, the positive vortex charge
is recovered \cite{Chen2}.

Also, since most systems have impurities, these affect both the motion of the
quasiparticles through scattering and through pinning of the vortices. Therefore
it is necessary to study the effect of the impurities. Such a study
has been carried out in the case of the d-density wave state \cite{Zhu}.
It is clear that, at the very least, the effect of the addition of the impurities is to
locally change the chemical potential.

In this work we consider several causes for disorder and their effect on the
vortex charge.
The effect of a charge inside a superconductor is in general screened. In usual
superconductors of the $s$-wave type, the dominant contribution is due to
Thomas-Fermi like screening, as in the normal state \cite{Fetter}. Due to the
presence of the gap the Fermi surface Friedel oscillations are suppressed. The
exponential screening acts on the scale of the Thomas-Fermi length which in general
is much smaller than the coherence length, and therefore any charge is very small.
However, in $d$-wave superconductors or in $s$-wave superconductors in the vicinity
of a vortex, the Friedel oscillations are important since in the first case there are
gapless states and in the second case there are states of an essentially normal
character in the vicinity of the vortex core \cite{Machida1}.
These oscillations act on a scale which is comparable to the coherence length, which
in type-II superconductors is small. Therefore in the vicinity of the vortex core
the screening effect, even though noticeable, does not change qualitatively the effect
of the charge depletion. This is shown in Fig. 1 of ref. \cite{Machida1} where the
charge oscillations near the core, even though depressed, are still visible and only
a quantitative change is observed. Therefore, for simplicity, we will neglect in this
work the effect of screening since the results will be qualitatively the same.
Also, as we will show later, the effect of the impurities has a local nature and only affects
significantly the physical quantities near the vortex core where screening has not fully
acted. Only far from the vortex core the screening of the Thomas-Fermi type will
strongly suppress the charge oscillations \cite{Machida1}.

\section{Vortices at low to intermediate fields}

Consider the lattice
formulation of a superconductor
in a magnetic field. Let us start from the Bogoliubov-de Gennes equations
${\cal H} \psi = \epsilon \psi$
where $\psi^{\dagger}(\mbf{r})=\left(u^*(\mbf{r}),
v^*(\mbf{r}) \right)$ and where the matrix Hamiltonian is given by
\begin{equation} \label{H}
{\cal H} = \left( \begin{array}{cc}
\hat{h} & \hat{\Delta} \\
\hat{\Delta}^{\dagger} & -\hat{h}^{\dagger} \\
\end{array} \right)
\end{equation}
with \cite{FT,PRB}
\begin{equation}\label{h}
\hat{h} = -t \sum_{\delta} e^{-\frac{ie}{\hbar c}
\int_{\mbf{r}}^{\mbf{r}+
\mbf{\delta}} \mbf{A}(\mbf{r}) \cdot
d\mbf{l}}
\hat{s}_{\delta} - \epsilon_F
\end{equation}
and
\begin{equation}\label{delta}
\hat{\Delta} = \Delta_0 \sum_{\delta} e^{\frac{i}{2}
\phi(\mbf{r})}
\hat{\eta}_{\delta} e^{\frac{i}{2} \phi(\mbf{r})}.
\end{equation}
The sums are over nearest neighbors ($\delta=\pm
\mbf{x}, \pm \mbf{y}$
on the square lattice); $\mbf{A}(\mbf{r})$ is the vector
potential associated with the uniform external magnetic field
$\mbf{B}=\nabla\times\mbf{A}$,
the operator $\hat{s}_{\delta}$ is defined through
its action on space dependent functions,
$\hat{s}_{\delta}u(\mbf{r}) = u(\mbf{r}+\delta)$, and
the operator $\hat{\eta}_{\delta}$ describes the symmetry
of the order parameter.
It is convenient to perform a singular gauge
transformation to eliminate the phase of the
off-diagonal term (\ref{delta}) in the matrix Hamiltonian.
We consider the unitary FT gauge transformation
${\cal H} \rightarrow {\cal U}^{-1} {\cal H} U$, where \cite{FT}
\begin{equation} \label{gauge}
{\cal U} = \left( \begin{array}{cc}
e^{i \phi_A(\mbf{r})} & 0 \\
0 & e^{-i \phi_B(\mbf{r})} \\
\end{array} \right)
\end{equation}
with $\phi_A(\mbf{r})+\phi_B(\mbf{r})=\phi(\mbf{r})$.
The phase field $\phi(\mbf{r})$ is decomposed at each site
of the two-dimensional lattice in two components $\phi_A(\mbf{r})$
and $\phi_B(\mbf{r})$ which are assigned respectively
to a set of vortices $A$, positioned at $\{\mbf{r}_i^A \}_{i=1,N_A}$,
and a set of vortices $B$, positioned at $\{\mbf{r}_i^B \}_{i=1,N_B}$.
The phase fields $\phi_{\mu=A,B}$ are defined through the equation
\begin{equation}
\nabla \times \nabla \phi_{\mu}
(\mbf{r}) = 2 \pi \mbf{z}
\sum_i \delta (\mbf{r}-\mbf{r}_i^{\mu})
\end{equation}
where the sum runs only over the $\mu$-type vortices.
After carying out the gauge transformation (\ref{gauge})
the Hamiltonian (\ref{H}) reads
%\begin{widetext}
\begin{equation}\label{Hp}
\mathcal{H}'=
\displaystyle
\left(
\begin{array}{cc}
-t\displaystyle\sum_{\delta}
e^{i\mathcal{V}^A_{\delta}\left(\mbf{r}\right)}\hat{s}_{\delta}
-\epsilon_F
&
\Delta_0\displaystyle\sum_{\delta}
e^{-i\frac{\delta\phi}{2}}
\hat\eta_{\delta}
e^{i\frac{\delta\phi}{2}}
\\
\Delta_0\displaystyle\sum_{\delta}
e^{-i\frac{\delta\phi}{2}}
\hat\eta^\dagger_{\delta}
e^{i\frac{\delta\phi}{2}}
&
t\displaystyle\sum_{\delta}
e^{-i\mathcal{V}^B_{\delta}\left(\mbf{r}\right)}\hat{s}_{\delta}
+\epsilon_F
\end{array}
\right).
\end{equation}
%\end{widetext}
The phase factors are given by \cite{PRB}
${\cal V}_{\delta}^{\mu}(\mbf{r})
=\int_{\mbf{r}}^{\mbf{r}+\delta}
\mbf{k}_s^{\mu}
\cdot d\mbf{l}$ and
$\delta\phi(\mbf{r})=\phi_A(\mbf{r})-\phi_B(\mbf{r})$,
where $\hbar\mbf{k}_s^{\mu} = m \mbf{v}_s^{\mu} =
\hbar \nabla
\phi_{\mu} -
\frac{e}{c} \mbf{A}$ is the superfluid momentum
vector for the $\mu$-supercurrent.
Physically, the vortices $A$ are only
visible to the particles and the vortices $B$ are only visible to the holes.
Each resulting $\mu$-subsystem is then in an effective magnetic field
\begin{equation}\label{Bmu}
\mbf{B}_{\mathrm{eff}}^\mu=-\frac{mc}{e} \nabla\times\mbf{v}^\mu_s
=\mbf{B}-\phi_0\mbf{z}\sum_i\delta^2(\mbf{r}-\mbf{r}_i^\mu)
\end{equation}
where each vortex carries an effective quantum magnetic flux $\phi_0$.
For the case of a regular vortex lattice \cite{FT,PRB}, these effective magnetic
fields vanish simultaneously on average if the magnetic unit cell
contains two vortices, one of each type.
More generally, in the absence of spatial symmetries, as it is the case
for disordered systems, these effective
magnetic fields $\mbf{B}_{\mathrm{eff}}^{\mu=A,B}$ vanish if
the numbers of vortices of the two types $A$ and $B$
are equal, i.e. $N_A=N_B$,
and their sum equals the number of elementary
quantum fluxes of the external
magnetic field penetrating the system.

The $\mu$-superfluid wave vector $\mbf{k}_s^{\mu}(\mbf{r})$
characterizes the supercurrents
induced by the $\mu$-vortices.
This
vector can be calculated for an arbitrary
configuration of vortices \cite{PRB} like
\be\label{kappamu}
\mbf{k}_s^{\mu}(\mbf{r})=2 \pi \int
\frac{d^2k}{(2 \pi)^2}
\frac{i \mbf{k} \times \mbf{z}}{ k^2+\lambda^{-2}}
\sum_{i=1}^\infty e^{i\mbf{k} \cdot (\mbf{r}-
\mbf{r}_i^{\mu})}.
\ee
As we take the London limit, which is valid for low
magnetic field and over most of the $H-T$ phase diagram in extreme
type-II superconductors such as cuprates, we assume that the size
of the vortex core is negligible and place each vortex core at
the center of a plaquette.

For the conventional $s$-wave case the operator characterizing
the symmetry of the order parameter is constant
$\hat\eta_{\mbf{\delta}}=\frac 14$ and the off-diagonal terms
of the Hamiltonian (\ref{Hp})
are then considerably simplified
\begin{equation}\label{Hs}
\mathcal{H}'=
\displaystyle
\left(
\begin{array}{cc}
-t\displaystyle\sum_{\delta}
e^{i\mathcal{V}^A_{\delta}\left(\mbf{r}\right)}\hat{s}_{\delta}
-\epsilon_F
&
\Delta_0
\\
\Delta_0
&
t\displaystyle\sum_{\delta}
e^{-i\mathcal{V}^B_{\delta}\left(\mbf{r}\right)}\hat{s}_{\delta}
+\epsilon_F
\end{array}
\right).
\end{equation}
Note that in this case the phase of the off-diagonal term is eliminated.

For the unconventional $d$-wave case the operator $\hat\eta_{\delta}$
takes the form $\hat\eta_{\delta}=(-1)^{\delta_y}\hat{s}_{\delta}$.
With these definitions the $d$-wave Hamiltonian can
be derived from the Hamiltonian (\ref{Hp}) and reads
\begin{equation}\label{Hd}
\mathcal{H}'=
\displaystyle
\left(
\begin{array}{cc}
-t\displaystyle\sum_{\delta}
e^{i\mathcal{V}^A_{\delta}\left(\mbf{r}\right)}\hat{s}_{\delta}
-\epsilon_F
&
\Delta_0\displaystyle\sum_{\delta}
e^{i\mathcal{A}_{\delta}(\mbf{r})+i\pi\delta_y}
\hat{s}_{\delta}
\\
\Delta_0\displaystyle\sum_{\delta}
e^{-i\mathcal{A}_{\delta}(\mbf{r})-i\pi\delta_y}
\hat{s}_{\delta}
&
t\displaystyle\sum_{\delta}
e^{-i\mathcal{V}^B_{\delta}\left(\mbf{r}\right)}\hat{s}_{\delta}
+\epsilon_F
\end{array}
\right)
\end{equation}
where the phase factor $\mathcal{A}_{\delta}(\mbf{r})$ has the form
\begin{eqnarray}
\mathcal{A}_{\delta}(\mbf{r})&=&\frac{1}{2}
\int_{\mbf{r}}^{\mbf{r}+\delta}\nonumber
\left(\nabla\phi_A-\nabla\phi_B\right)\cdot d\mbf{l}\\
&=&\frac{1}{2}\label{Adelta}
\int_{\mbf{r}}^{\mbf{r}+\delta}
\left(\mbf{k}_s^A-\mbf{k}_s^B\right)\cdot d\mbf{l}.
\end{eqnarray}
In the Hamiltonian (\ref{Hd}) and in Eq. \ref{Adelta} the vector
\begin{equation}
\mbf{a}_s=\frac 12\left(\mbf{k}_s^A-\mbf{k}_s^B\right)
\end{equation}
acts as an internal gauge field independent of the external magnetic
field \cite{PRB}. The associated internal magnetic field
$\mbf{b}=\nabla\times\mbf{a}_s$ consists of opposite
$A-B$ spikes fluxes carying each one half of the magnetic quantum flux
$\phi_0$, centered in the vortex cores and vanishing on average
since the numbers of $A$- and $B$-type vortices are the same.

The solution of the BdG equations gives the spectrum and the wave functions
of the quasiparticles.
As the effective magnetic fields experienced by the
particles and the holes vanish on average, within the gauge transformation
we are allowed to use periodic boundary conditions
on the square lattice ($\Psi(x+nL,y+mL)=\Psi(x,y)$ with $n,m\in
\Bbb{Z}$). The $L\times L$ original lattice becomes then a
magnetic supercell where the impurities are placed at random and
where the vortices are placed in such a way as to minimize their
total energy. The disorder induced by the
impurities in the system is then established over a length $L$.
Thus in order to compute the eigenvalues and eigenvectors of the
Hamiltonian we seek for eigensolutions in the
Bloch form
$\Psi^\dagger_{n\mbf{k}}(\mbf{r})=e^{-i\mbf{k}\cdot\mbf{r}}
(U^*_{n\mbf{k}},V^*_{n\mbf{k}})$ where $\mbf{k}$ is a point of the
Brillouin zone. We diagonalize then the Hamiltonian
$e^{-i\mbf{k}\cdot\mbf{r}}\mathcal{H}e^{i\mbf{k}\cdot\mbf{r}}$ for
a large number of points $\mbf{k}$ in the Brillouin zone and for
many different realizations (around 100) of the random impurity
positions and of the correlated vortex positions.

The density of states (DOS) and the local density of states (LDOS) for the cases
of vortex disorder \cite{Lages1} and the combined effects of vortex disorder
and impurity scattering \cite{Lages2} were studied recently. The disorder
in general increases the DOS at low energies, by filling the gap in the s-wave
case and originating a finite density of states at zero energy in the
d-wave case. An approximate scaling regime was obtained in the last case.
Also, the results for the LDOS are in qualitative agreement with STM experiments
if most vortices are pinned at the impurity locations.

The electron density is calculated in the usual
way
\be
n(\vec{r}) = 2 \sum_{i} \left( |u_{i}(\vec{r})|^2 f(E_{i}) +
|v_{i}(\vec{r})|^2 (1-f(E_{i}) \right)
\ee
where $E_{i}$ are the energy eigenvalues and $f(E_{i})$ is the Fermi
function.

\subsection{Vortex lattice}

\begin{figure}
\includegraphics[width=0.8\columnwidth]{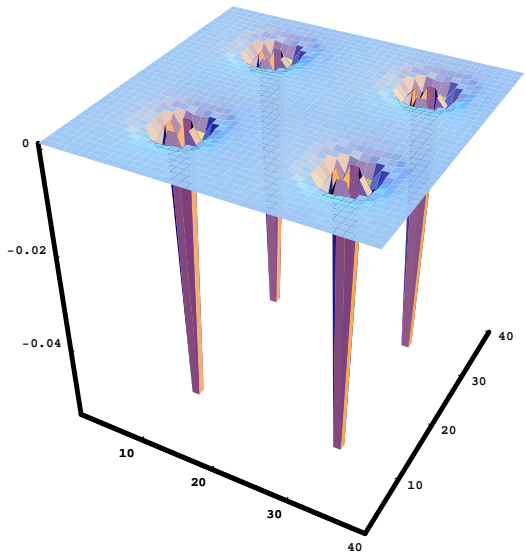}
\caption{\label{fig1} (color online) Electron density
$n(\mathbf{r})-n_{\mathrm{bulk}}$ for s-wave symmetry in a regular
vortex lattice. Here $\Delta=t$, $\mu=-2.2t$ and
$n_{\mathrm{bulk}}=0.377$. }
\end{figure}

\begin{figure}
\includegraphics[width=0.8\columnwidth]{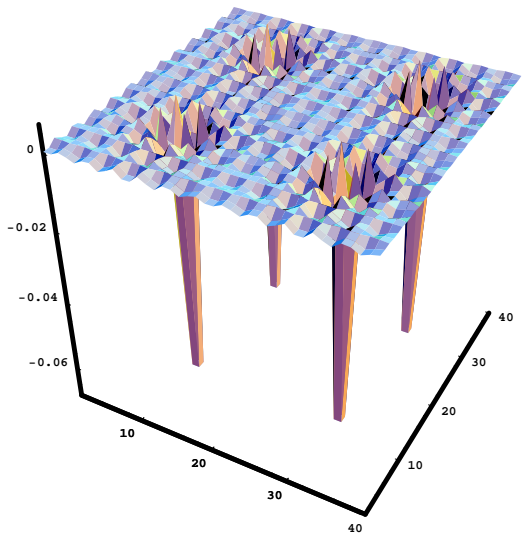}
\caption{\label{fig2} (color online) Electron density
$n(\mathbf{r})-n_{\mathrm{bulk}}$ for d-wave symmetry in a regular
vortex lattice. Here $\Delta=t$, $\mu=-2.2t$ and
$n_{\mathrm{bulk}}=0.509$. }
\end{figure}

\begin{figure}
\includegraphics[width=0.8\columnwidth]{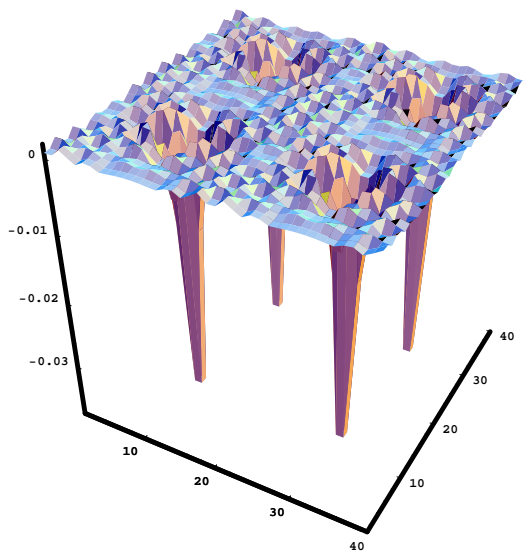}
\caption{\label{fig3} (color online) Electron density
$n(\mathbf{r})-n_{\mathrm{bulk}}$ for d-wave symmetry in a regular
vortex lattice. Here $\Delta=0.25t$, $\mu=-2.2t$ and
$n_{\mathrm{bulk}}=0.36$. }
\end{figure}

We consider first, for completeness, a regular distribution of vortices both for s-wave
and d-wave pairings. At not very high fields, the vortices are sufficiently
apart and the vortex charge is depleted from the vortex cores, as mentioned above.
Therefore the vortices are positively charged. In Fig. \ref{fig1} we consider the s-wave
pairing case and in Fig. \ref{fig2} we consider the d-wave pairing case. In this last case
a checkerboard modulation of the charge density is seen. Both sets of
results are presented for the case $\Delta=t$.
Linearizing the BdG equations defined in a continuum close to the nodes leads naturally to the
definition of two velocities the Fermi velocity, $v_F$, and a velocity
$v_{\Delta}=\Delta_0/p_F$, where $p_F$ is the Fermi momentum. This velocity
denotes the slope of the gap at the node \cite{Lee,FT}. Defining the anisotropy
of the Dirac cone as the ratio $\alpha_D=v_F/v_{\Delta}$ which in the
lattice case translates to $\alpha_D=t/\Delta_0$, the case considered above
$\Delta=t$ is called the isotropic case.
In many d-wave superconductors the anisotropy is actually stronger. We present for
comparison in Fig. \ref{fig3} results for d-wave pairing considering
$\Delta=0.25t$ (note that in high-$T_c$ materials the anisotropy is actually stronger
of the order of $t=15\Delta$). As the anisotropy increases the charge depletion
decreases in depth but extends in area.

\begin{figure}
\includegraphics[width=0.8\columnwidth]{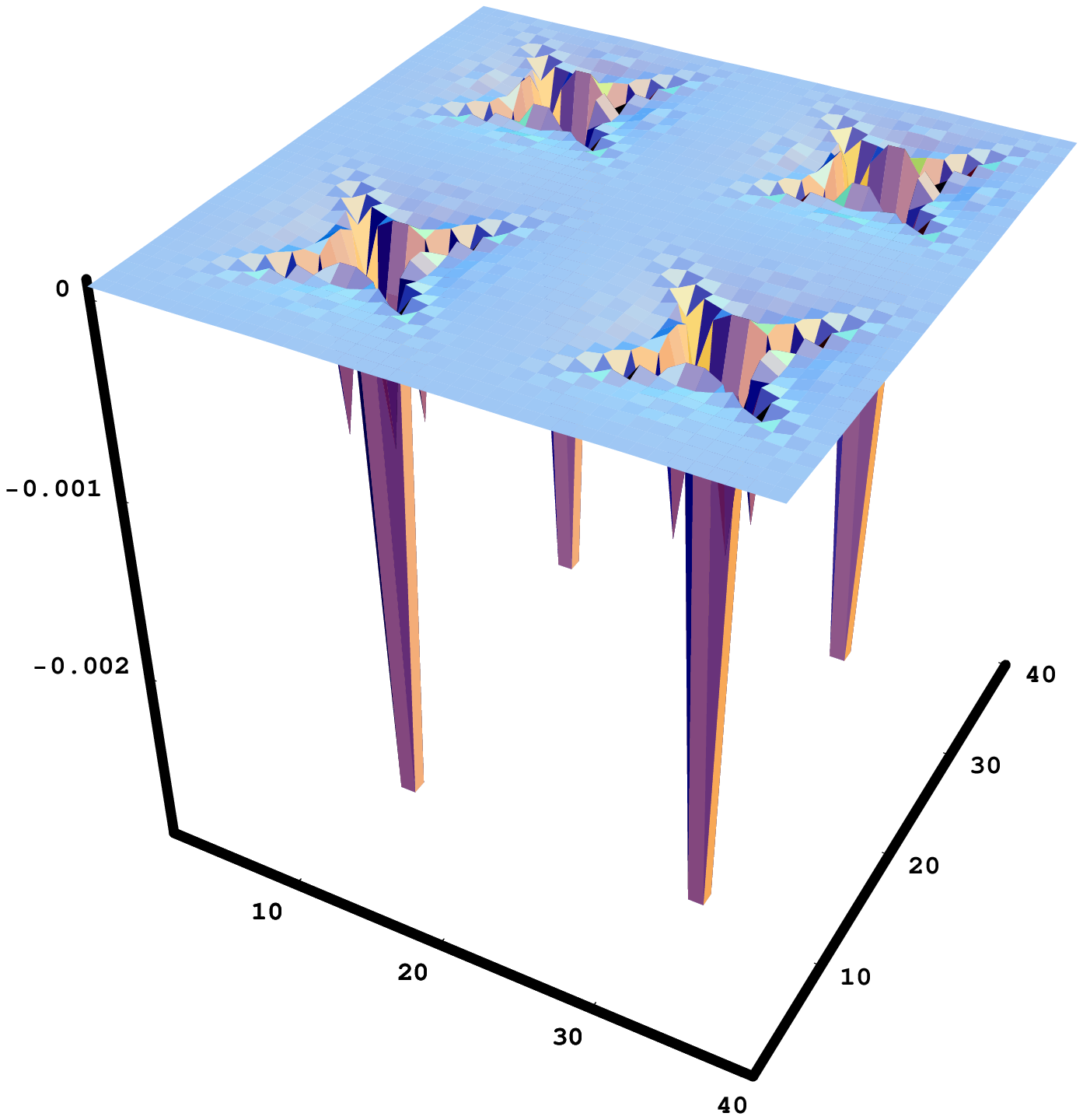}
\caption{\label{fig4} (color online) Electron density
$n(\mathbf{r})-n_{\mathrm{bulk}}$ for s-wave symmetry in a regular
vortex lattice. Here $\Delta=t$, $\mu=-0.1t$ and
$n_{\mathrm{bulk}}=0.965$. }
\end{figure}

%\begin{figure}
%\includegraphics[width=0.8\columnwidth]{mun0.1d.eps}
%\caption{\label{fig5} (color online) Electron density
%$n(\mathbf{r})-n_{\mathrm{bulk}}$ for d-wave symmetry in a regular
%vortex lattice. Here $\Delta=t$, $\mu=-0.1t$ and
%$n_{\mathrm{bulk}}=0.977$. }
%\end{figure}

\begin{figure}
\includegraphics[width=0.8\columnwidth]{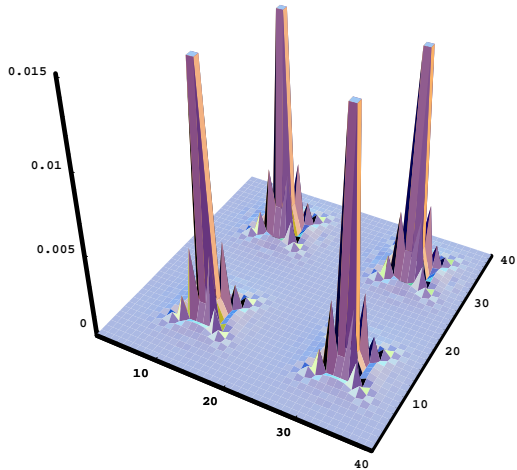}
\caption{\label{fig6} (color online) Electron density
$n(\mathbf{r})-n_{\mathrm{bulk}}$ for s-wave symmetry in a regular
vortex lattice. Here $\Delta=t$, $\mu=0.5t$ and
$n_{\mathrm{bulk}}=1.17$. }
\end{figure}

%\begin{figure}
%\includegraphics[width=0.8\columnwidth]{mup0.5d.eps}
%\caption{\label{fig7} (color online) Electron density
%$n(\mathbf{r})-n_{\mathrm{bulk}}$ for d-wave symmetry in a regular
%vortex lattice. Here $\Delta=t$, $\mu=0.5t$ and
%$n_{\mathrm{bulk}}=1.11$. }
%\end{figure}

We have considered a low density regime. We can vary the chemical potential
(and the band-filling). In Fig. \ref{fig4} we present the electron density
for a situation close to half-filling for the s-wave case.
In Fig. \ref{fig6}
we consider a case larger than half-filling.
In this case there is a charge accumulation (or hole depletion).
Indeed as
explained before \cite{Hayashi} the density of the dominant carriers is depleted near
the vortex core. Also, we have checked explicitly that at half-filling the vortex
charge vanishes \cite{Hayashi} and the electron density is uniform.
The results for the d-wave case show the same trend.

\subsection{Vortex disorder}

We consider now the case when the vortices are distributed randomly due
to some strong pinning effects, but neglect the effect of impurities on
the motion of the quasiparticles. Therefore, the superfluid velocities
are determined by a random distribution of the locations of the vortices,
assumed static.

\begin{figure}
\includegraphics[width=0.8\columnwidth]{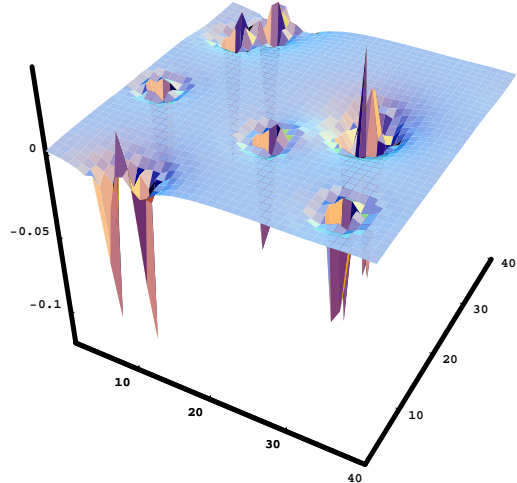}
\caption{\label{fig8} (color online) Electron density
$n(\mathbf{r})-n_{\mathrm{bulk}}$ for s-wave symmetry in a
disordered vortex lattice. Here the magnetic field is $B=1/160$
which corresponds to 10 vortices in a 40$\times$40 unit cell. Here
$\Delta=t$, $\mu=-2.2t$ and $n_{\mathrm{bulk}}=0.375$. }
\end{figure}

\begin{figure}
\includegraphics[width=0.8\columnwidth]{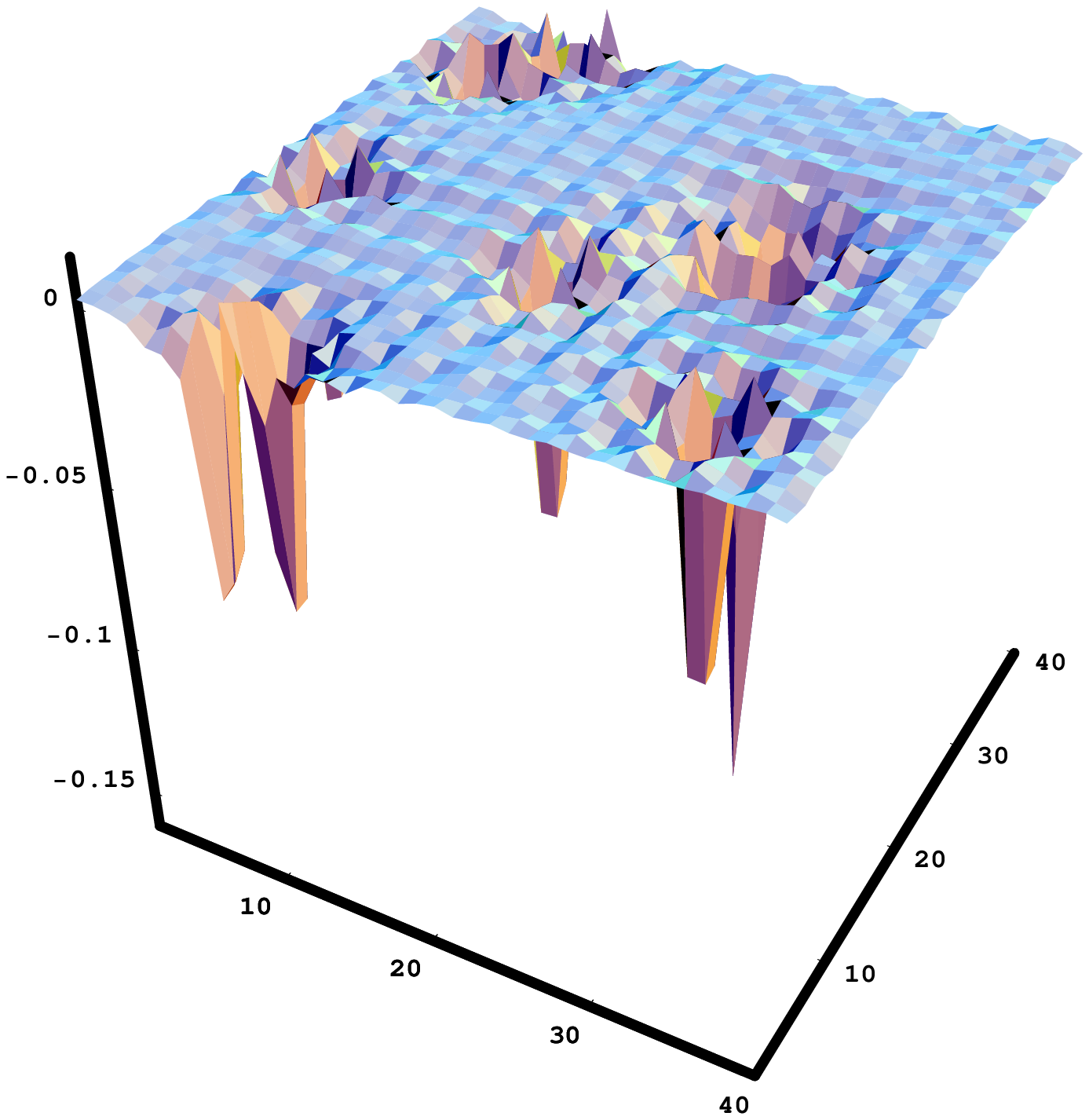}
\caption{\label{fig9} (color online) Electron density
$n(\mathbf{r})-n_{\mathrm{bulk}}$ for d-wave symmetry in a
disordered vortex lattice. Here the magnetic field is $B=1/160$
which corresponds to 10 vortices in a 40$\times$40 unit cell. Here
$\Delta=t$, $\mu=-2.2t$ and $n_{\mathrm{bulk}}=0.507$. }
\end{figure}

In Fig. \ref{fig8} we consider the case of s-wave pairing and in Fig. \ref{fig9} we consider
d-wave pairing for a particular random distribution of the vortices (note
that there is no sum over random configurations, like in the calculation
of the LDOS).
When a vortex is isolated the distortion of the charge
profile is very similar to the lattice case. However, if two vortices are
pinned nearby, the charge profile is significantly changed. In conjunction
with the charge depletion, there is a sign reversal of the electron
density. This is particularly visible in the s-wave case but also occurs
for the d-wave symmetry. In the region where the two vortex cores are located
there is a charge accumulation which leads to a local negative charge
with respect to the bulk value. These fluctuations are of a similar order
of magnitude as the charge depletion at the single vortices.
However, the integration of the charge density in the neighborhood of the
two vortices is still positive, despite the strong negative oscillations.
The total charge in the
case when two vortices are close by, even though still positive, is considerably smaller.

\begin{figure}
\includegraphics[width=0.8\columnwidth]{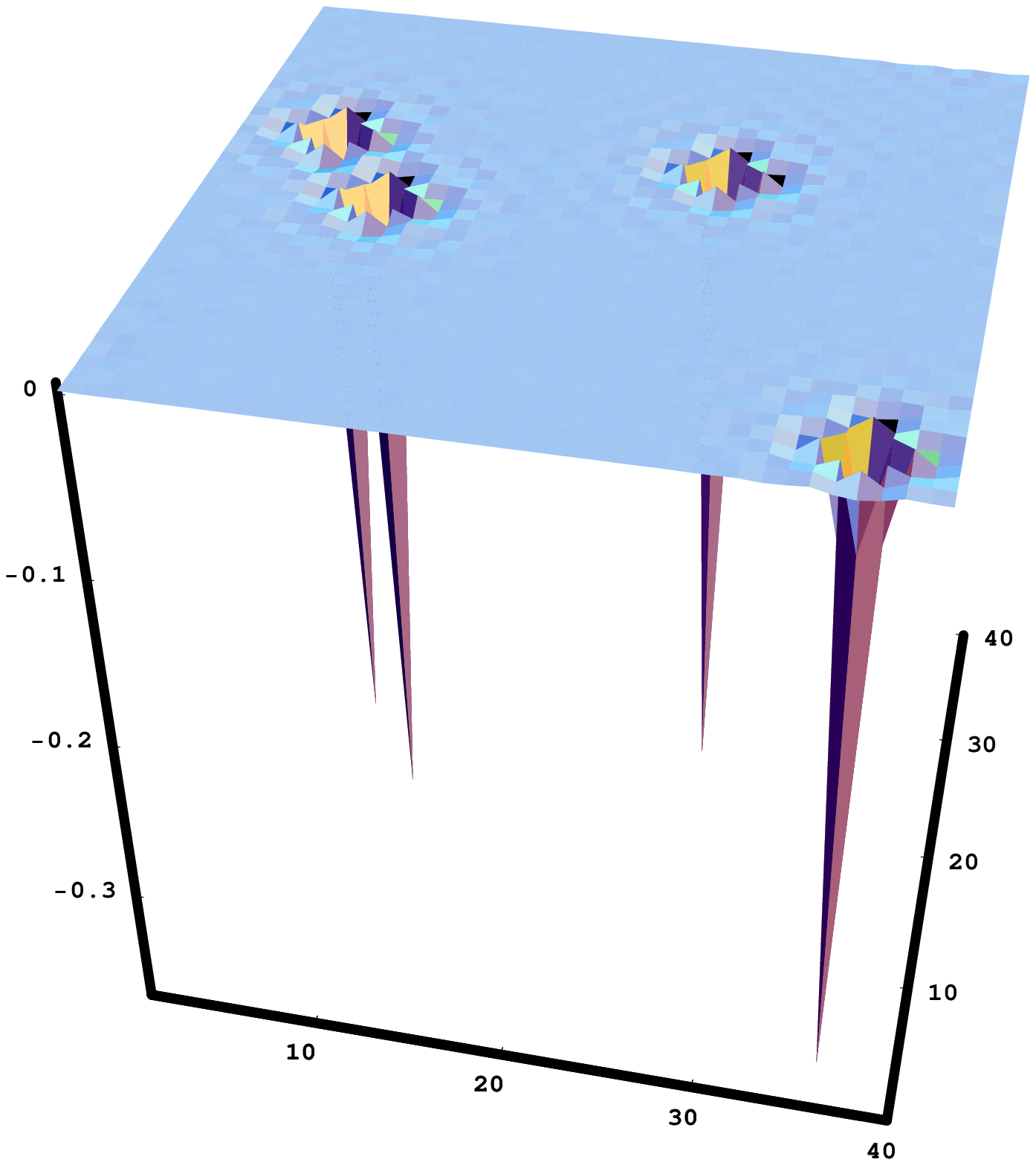}
\caption{\label{fig12} (color online) Electron density
$n(\mathbf{r})-n_{\mathrm{bulk}}$ for d-wave symmetry with
impurities and no vortices. Here $\Delta=t$, $\mu=-2.2t$, $U=5t$ and
$n_{\mathrm{bulk}}=0.509$. }
\end{figure}

\begin{figure}
\includegraphics[width=0.8\columnwidth]{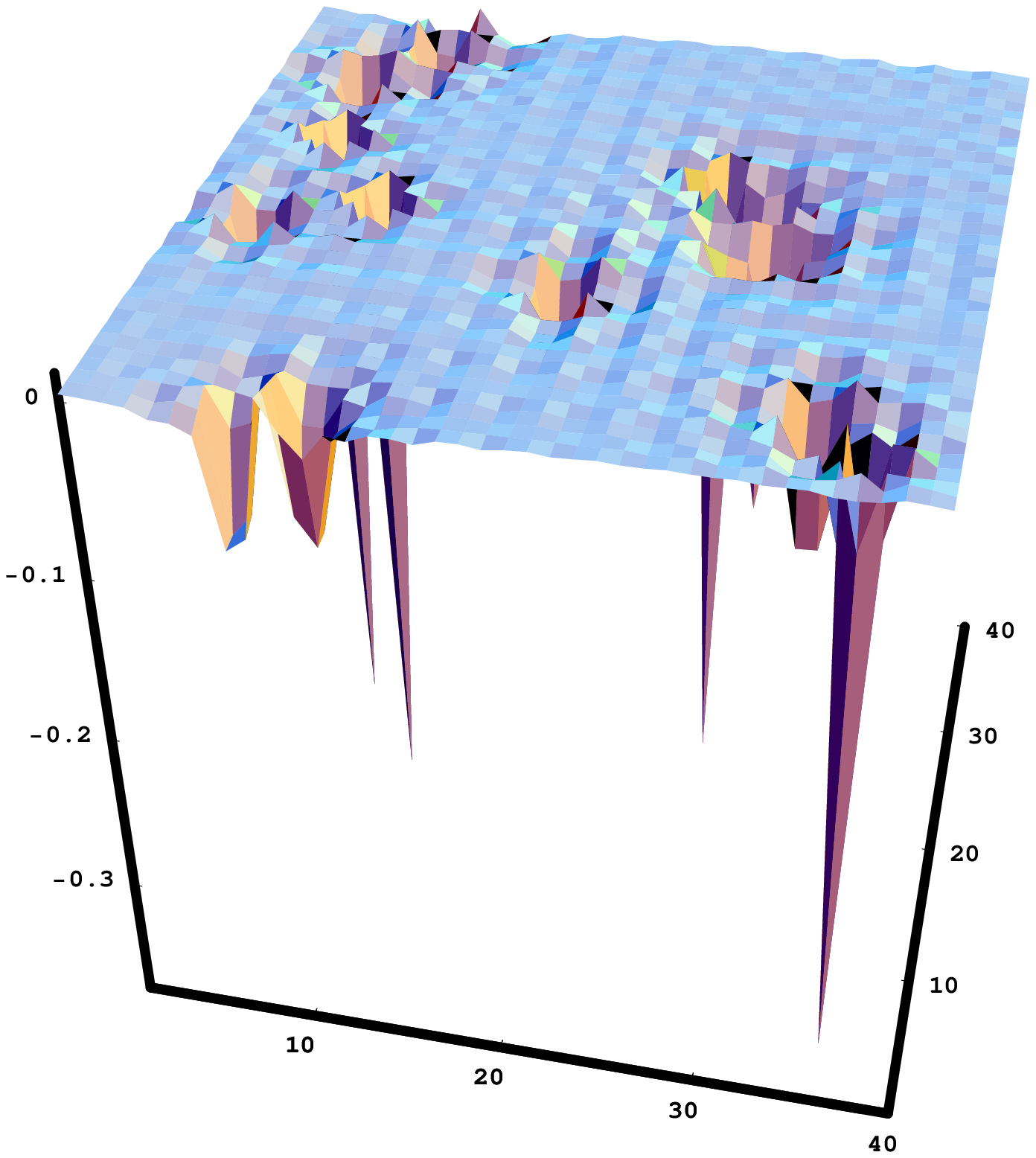}
\caption{\label{fig13} (color online) Electron density
$n(\mathbf{r})-n_{\mathrm{bulk}}$ for d-wave symmetry in a
disordered vortex lattice with impurities. Here $\Delta=t$,
$\mu=-2.2t$, $U=5t$, $B=1/160$ and $n_{\mathrm{bulk}}=0.505$. }
\end{figure}

\begin{figure}
\includegraphics[width=0.8\columnwidth]{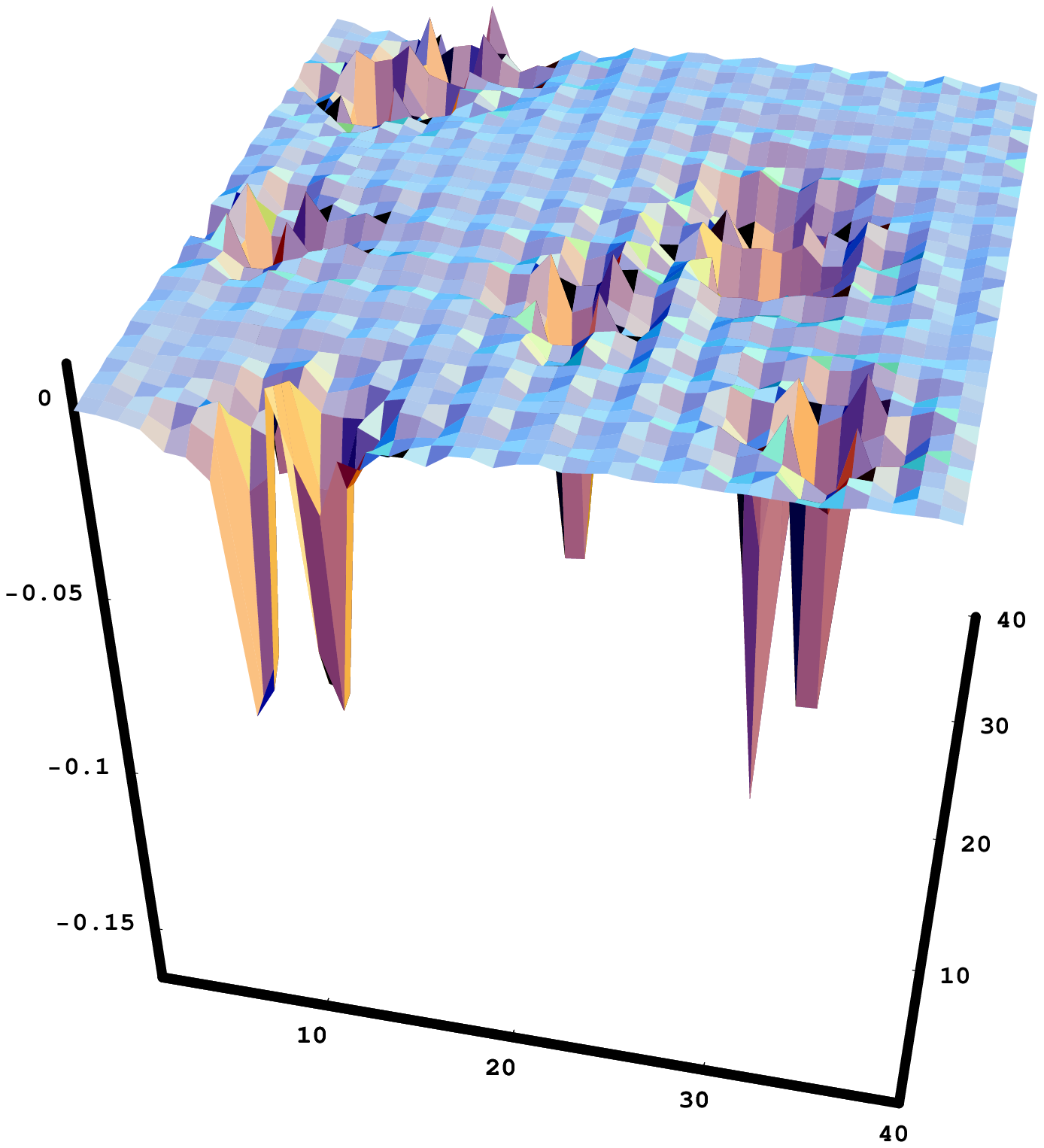}
\caption{\label{fig14}
(color online) Difference in the electron density between the cases with and
without vortices shown in Figs. \ref{fig12} and \ref{fig13}.
}
\end{figure}

\begin{figure}
\includegraphics[width=0.8\columnwidth]{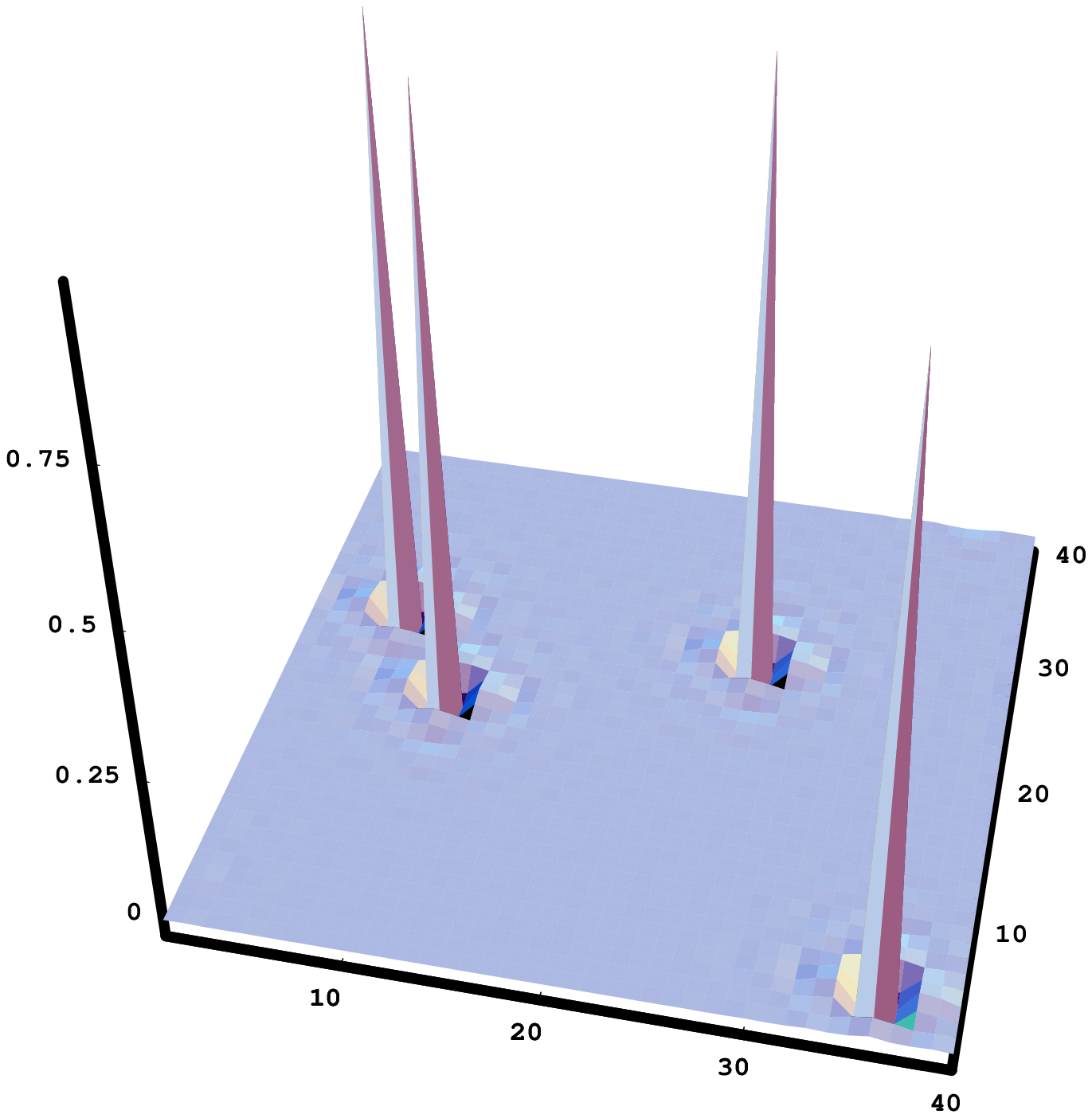}
\caption{\label{fig15} (color online) Electron density
$n(\mathbf{r})-n_{\mathrm{bulk}}$ for d-wave symmetry with
impurities and no vortices. Here $\Delta=t$, $\mu=-2.2t$, $U=-5t$
and $n_{\mathrm{bulk}}=0.513$. }
\end{figure}

\begin{figure}
\includegraphics[width=0.8\columnwidth]{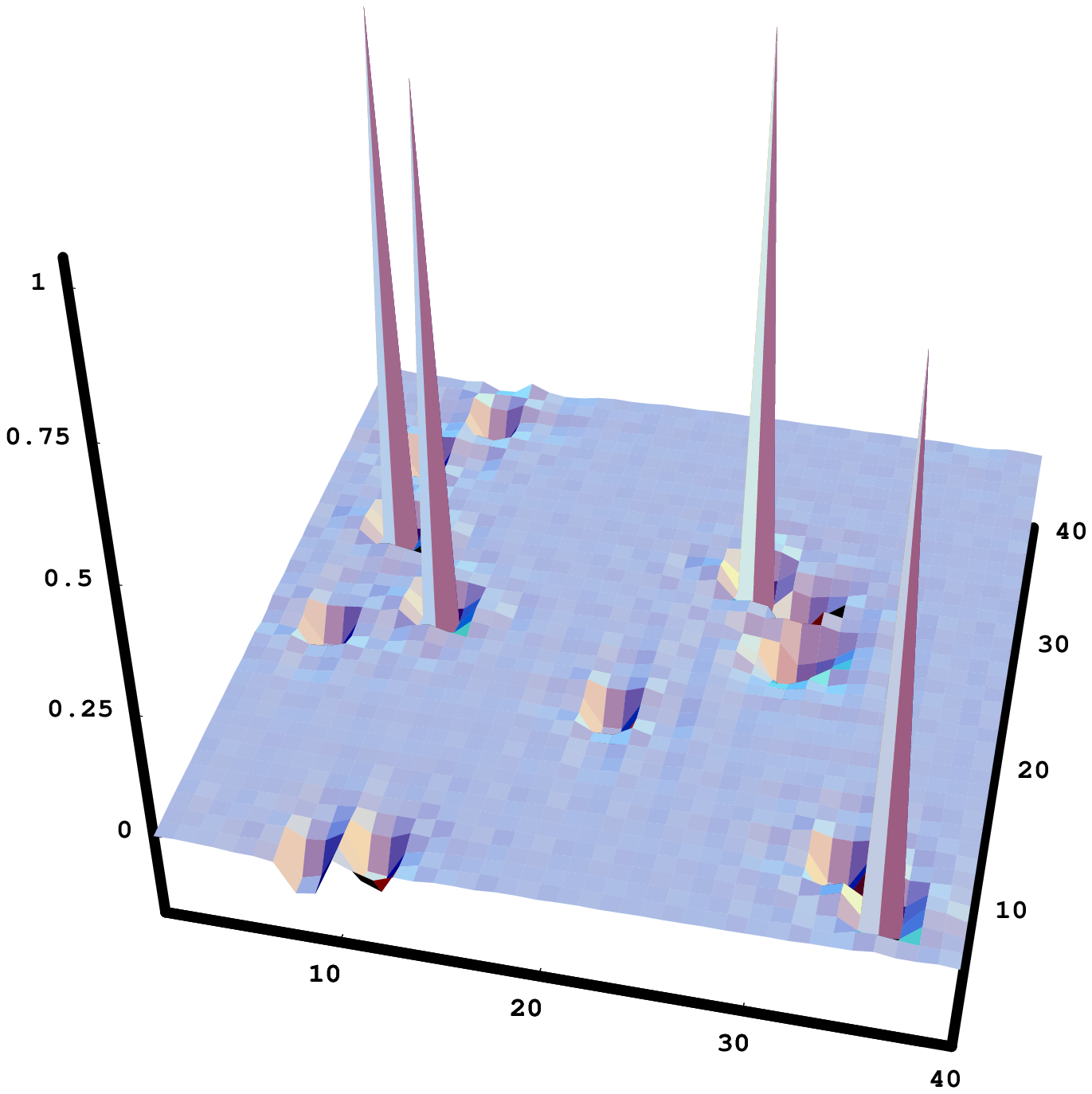}
\caption{\label{fig16} (color online) Electron density
$n(\mathbf{r})-n_{\mathrm{bulk}}$ for d-wave symmetry in a
disordered vortex lattice with impurities. Here $\Delta=t$,
$\mu=-2.2t$, $U=-5t$, $B=1/160$ and $n_{\mathrm{bulk}}=0.509$. }
\end{figure}

\begin{figure}
\includegraphics[width=0.8\columnwidth]{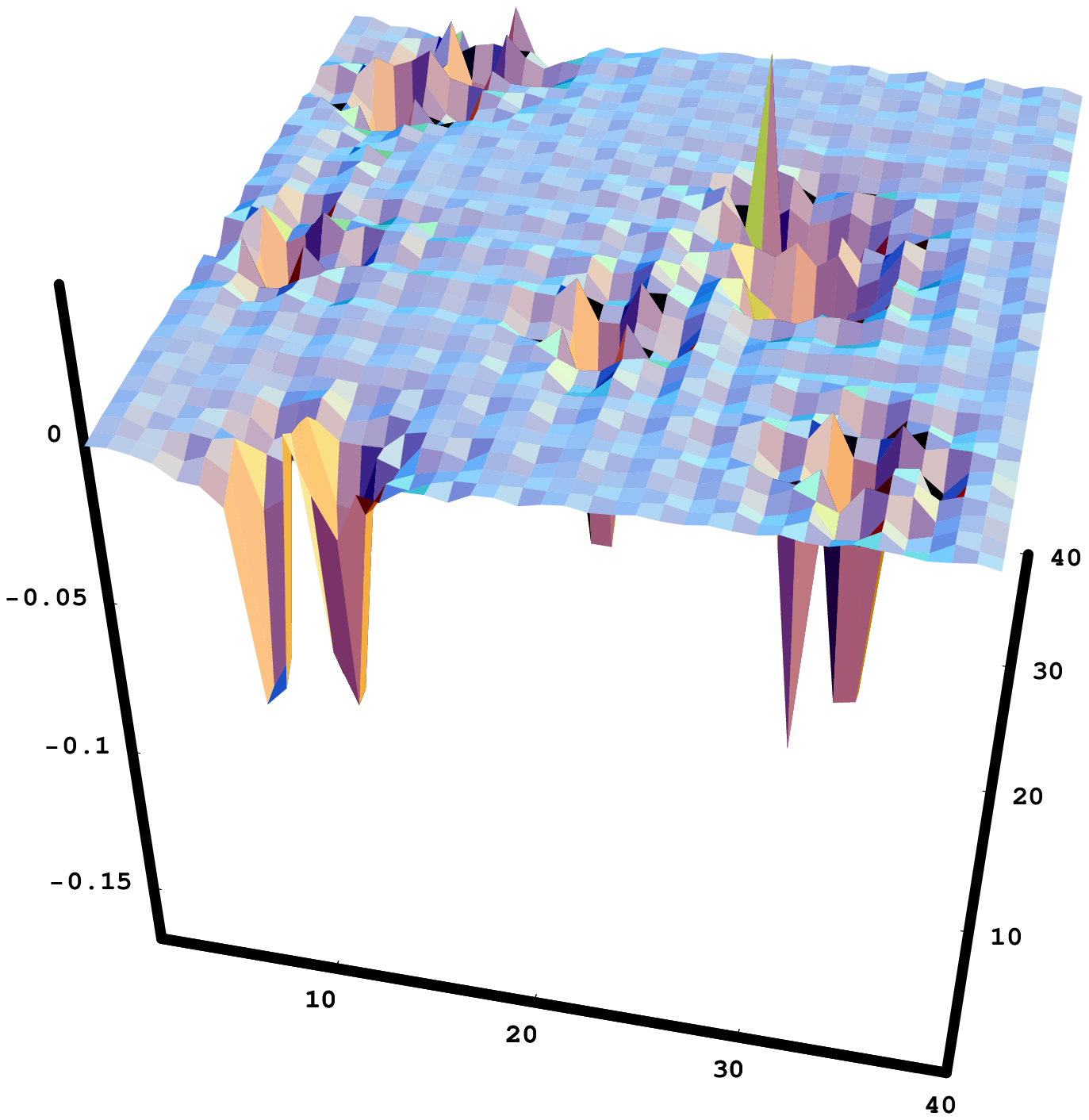}
\caption{\label{fig17}
(color online) Difference in the electron density between the cases with and
without vortices shown in Figs. \ref{fig15} and \ref{fig16}.
}
\end{figure}

\subsection{Effect of impurities}

We consider now the effect of impurities. These are introduced
at the Hamiltonian level, with the substitution
\[
{\hat h} \rightarrow {\hat h} + \mathcal{U}(\mathbf{r})
\]
Here $\mathcal{U}(\mathbf{r})$ is the potential due to the impurities
placed randomly in the system.
We model the disorder using the
binary alloy model \cite{Atkinson}.
At each impurity site it costs an energy $U$ to place an
electron (it acts as a local shift on the chemical potential).
The impurities are randomly distributed over a $L\times L$ periodic
two-dimensional lattice and play the role of pinning centers for
the vortices.
In general, it is favorable that a vortex
is located in the vicinity of an impurity \cite{pinning}.
However, in this work we will be considering random distributions of the
vortices and impurities and study the electron density throughout the system
with an arbitrary distribution of the vortices and impurities.
Due to the non-homogeneous nature of the order parameter the BdG equations
have to be solved self-consistently.

The effect of the impurities is the expected one. If the potential is repulsive
the electron density is lowered and if the potential is attractive the electron
density is increased leading to an electron accumulation and a sign reversal.
There are sharp
peaks at the impurity locations that mask the effect of the vortex charge
in the material. If a vortex is pinned at an impurity site (as in most systems
they are) the effect of the impurity potential is quite strong. For moderate
values of the potential the effect is smaller but still noticeable.

%\begin{figure}
%\includegraphics[width=0.8\columnwidth]{up100_d.eps}
%\caption{\label{fig10} (color online) Electron density
%$n(\mathbf{r})-n_{\mathrm{bulk}}$ for d-wave symmetry in a
%disordered vortex lattice with impurities ($U>0$). Here the magnetic
%field is $B=1/160$ which corresponds to 10 vortices in a
%40$\times$40 unit cell. Here $\Delta=t$, $\mu=-2.2t$, $U=100t$ and
%$n_{\mathrm{bulk}}=0.504$. }
%\end{figure}

%\begin{figure}
%\includegraphics[width=0.8\columnwidth]{um100_d.eps}
%\caption{\label{fig11} (color online) Electron density
%$n(\mathbf{r})-n_{\mathrm{bulk}}$ for d-wave symmetry in a
%disordered vortex lattice with impurities ($U<0$). Here the magnetic
%field is $B=1/160$ which corresponds to 10 vortices in a
%40$\times$40 unit cell. Here $\Delta=t$, $\mu=-2.2t$, $U=-100t$ and
%$n_{\mathrm{bulk}}=0.509$. }
%\end{figure}

To compare the effects of the impurities and the vortices, we show in Figs. \ref{fig12}
and \ref{fig13} the electron density for $U=5t$ without and with vortices for a
specific distribution
of the impurities. In Fig. \ref{fig14} we plot the difference between the two cases. We see
that even though the impurities have a strong (local) influence on the charge
distribution, the contributions from the impurities and the vortices add up and the
difference is not negligible. The same is observed for the attractive case $U=-5t$
shown in Figs. \ref{fig15}, \ref{fig16} and \ref{fig17}.
Also, note that increasing the impurity potential for values $|U|>5t$
does not change qualitatively the charge since $|U|=5t$ is already a large value.

Clearly, when the vortices are diluted and neglecting the effect of the
impurities, the interaction between the vortices
is not so important and the behavior of the system is not very different
from an isolated vortex. A possible exception is the regime of very high
magnetic fields. But for low to intermediate fields, as seen from Figs. \ref{fig1}
and \ref{fig2},
the distortion of the electron density occurs close to the vortex locations
and the behavior is characteristic of a single vortex. The vortex charge has
been studied for a single vortex, as mentioned above, considering a vortex
with a quantum of flux. It is therefore also interesting to consider the influence
of an impurity in the single vortex case.

\section{Single vortex}

We consider then the effect of an impurity in a s-wave vortex.
We solve the Bogoliubov-de Gennes equations on a continuum \cite{Gygi} introducing
an impurity as a disc of small radius $d$ centered at the vortex location.

The general solution is obtained solving numerically the
BdG equations.
In the s-wave case these can be written
as
\bea
\left( H_e + U(\mbf{r}) \right) u(\mbf{r}) + \Delta(\mbf{r}) v(\mbf{r}) &=& \epsilon u(\mbf{r})
\nonumber \\
- \left( H_e^* + U(\mbf{r}) \right) v(\mbf{r}) + \Delta^*(\mbf{r}) u(\mbf{r})
&=& \epsilon v(\mbf{r})
\eea
Here
\[ H_e (\mbf{r}) = \frac{1}{2m} \left( \frac{\hbar}{i} \mbf{\nabla} - \frac{e}{c} \mbf{A}
\right)^2 + U(\mbf{r}) -E_F
\]
where $m$ is the electron mass, $\mbf{A}$ is the vector potential,
$E_F$ the Fermi energy and $U(\mbf{r})$ is the potential originated from
the impurity.

\begin{figure}
\includegraphics[width=6cm,height=7cm]{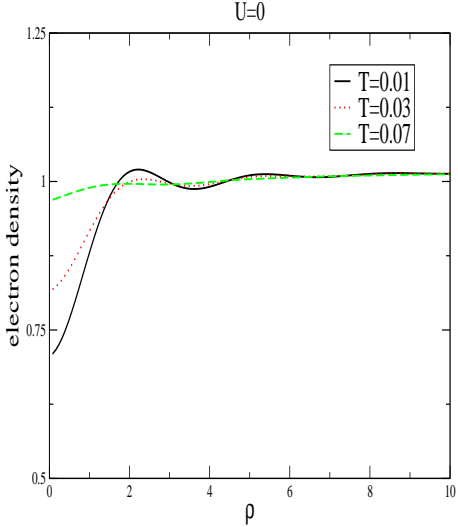}
\caption{\label{fig18}
(color online)
Electron density of a single vortex as a function of distance for different values of $T$.
The parameters are chosen as $E_F=1, V=1.6, R=80$. The energies
are in units of the Fermi energy and the lengths in units of the inverse
of the Fermi momentum.
}
\end{figure}

The BdG equations are solved using the decompositions \cite{Gygi}:
\bea
u^i(\rho,\varphi) &=& \sum_{\mu,j} c^i_{\mu,j} e^{i \varphi \mu}
\phi_{j,\mu}(\rho) \nonumber \\
v^i(\rho,\varphi) &=& \sum_{\mu,j} d^i_{\mu,j} e^{i \varphi \mu}
\phi_{j,\mu}(\rho)
\eea
and choosing a gauge such that $\Delta(\rho,\varphi)=\Delta(\rho) e^{-im\varphi}$.
appropriate for a vortex containing $m$ flux quanta.
Here $\mu$ is an integer,
$j$ is the number of the zero of the Bessel functions
in a disc of radius $R$ and the normalized functions $\phi_{j,\mu}$ constitute a complete
set over the zeros and are defined by
\be
\phi_{j,\mu} = \frac{\sqrt{2}}{R J_{\mu+1}(\alpha_{j,\mu})} J_{\mu} \left( \alpha_{j,\mu}
\frac{\rho}{R} \right)
\ee
Here $J_{\mu}$ is the Bessel function of order $\mu$ and $\alpha_{j,\mu}$ is the $j^{th}$ zero
of the Bessel function $J_{\mu}$. The functions $\phi_{j,\mu}$, by construction, are zero
at the border of the disc $\rho=R$.
In the basis of the functions $\phi_{j,\mu}$ we have to solve a set of equations for
the $c$ and $d$ coefficients of the form
\[
\sum_{\mu^{\prime},j^{\prime}}
\left( \begin{array}{cc}
T_{\mu,j;\mu^{\prime},j^{\prime}}^- & \Delta_{\mu,j;\mu^{\prime},j^{\prime}} \\
\Delta_{\mu,j;\mu^{\prime},j^{\prime}}^T & T_{\mu,j;\mu^{\prime},j^{\prime}}^+ \\
\end{array} \right)
\left( \begin{array}{c}
c_{\mu^{\prime},j^{\prime}} \\
d_{\mu^{\prime},j^{\prime}} \\
\end{array} \right)
= E
\left( \begin{array}{c}
c_{\mu,j} \\
d_{\mu,j} \\
\end{array} \right)
\]
Here the various components are diagonal in the angular momentum
$T_{\mu,j;\mu^{\prime},j^{\prime}}^{\pm}=\delta_{\mu, \mu^{\prime}}
T_{\mu,;j,j^{\prime}}^{\pm}$ and the same for the off-diagonal
terms. In the case of strongly type-II superconductors and if we are
interested in the low energy states, which are particularly relevant at small
distances from the vortex core, the vector potential may be neglected.
The various terms are then given by
\bea
& & T_{\mu;j,j^{\prime}}^{\pm} = \mp \left(\frac{\hbar^2}{2m} \left( \frac{\alpha_{j,\mu
}}{R} \right)^2 -E_F \right) \delta_{j,j^{\prime}} \nonumber \\
&\mp & U \int_0^{d} d\rho \rho \phi_{j,\mu}(\rho)
\phi_{j^{\prime},\mu}(\rho) \nonumber
\eea
and
\be
\Delta_{\mu;j,j^{\prime}} = \int_0^R d \rho \rho \phi_{j,\mu}(\rho)
\Delta (\rho) \phi_{j^{\prime},\mu+m}(\rho)
\ee
where we have used that
\[
U(\rho) = U \theta (d-\rho)
\]
The gap function is obtained in the usual way
\be
\Delta (\rho) = V \sum_{\mu,i (|E_i| \leq \omega_D)} \bar{u}_{\mu,i}(\rho)
\bar{v}_{\mu,i}(\rho) \left( 1- 2 f(E_{\mu,i}) \right)
\ee
with
$u_{\mu,i}(\rho,\varphi)= e^{i\varphi \mu} \bar{u}_{\mu,i}(\rho)$
and
$v_{\mu,i}(\rho,\varphi)= e^{i\varphi \mu+m} \bar{v}_{\mu,i}(\rho)$ where $V$ is the attractive
effective interaction between the electrons.
The BdG equations are solved self-consistently, as mentioned above.

\begin{figure}
\includegraphics[width=0.8\columnwidth]{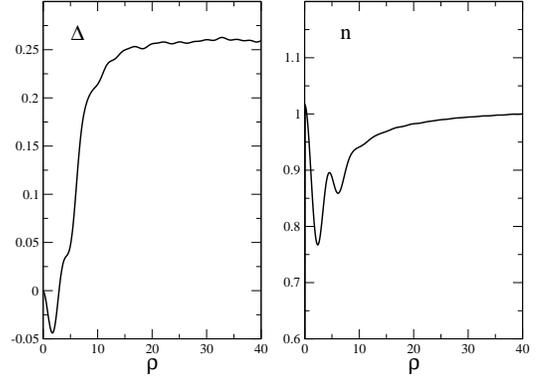}
\caption{\label{fig19}
Vortex with a double flux $2\Phi_0$.
Left panel: $\Delta$ as a function of distance.
Right panel: electron density as a function of distance
}
\end{figure}

Considering first the case of a vortex enclosing a quantum of flux ($m=1$)
we show, for completeness, in Fig. \ref{fig18} the charge profile close to the vortex
core, considering first $U=0$ (no impurity) \cite{Hayashi}. Close to the vortex core
the electron density decreases with respect to the bulk value.
The effect is specially evident at low $T$. As the temperature increases
the charge depletion decreases. This result is due to the Kramer and Pesch
effect: the vortex core size increases as the temperature increases and,
since the charge depletion may be related to the variation of the gap
function \cite{Jin}, this derivative decreases as $T$ increases and therefore the
charge depletion decreases as $T$ increases. As proposed in ref. \cite{Jin}
the electrostatic potential has contributions that are due to the difference
of the gap function at the vortex with respect to the bulk value, due to the
derivatives of the amplitude and of the phase of the gap function. Through
Poisson's equation these dependences carry to the electron density.
Note that at low $T$ there
are oscillations in the electron charge density. Also note that we are considering
here the full quantum limit where we have access to the vortex structure
inside the vortex core. In the previous sections the method neglected the vortex
core and we had no access to the true vortex core.

As we saw in section II.B, when two vortices are close by there are also
oscillations in the charge density that result, in this case, from the
vicinity of two vortices. On a large scale (where the vortex core is averaged out)
two vortices nearby may appear similar to a vortex containing two flux quanta
(note however that the energy of a double flux vortex is higher than two single
quantum flux vortices). In Fig. \ref{fig19} we consider the case of a vortex
containing two flux quanta ($m=2$). The inner structure of the vortex is somewhat
different \cite{96,Virtanen}. The gap function has a node near the origin and there are opposing
currents in the same regime, as also obtained using the Andreev Hamiltonian. In the right
panel of Fig. \ref{fig19} we show the electron density profile. As one approaches the core
the charge is depleted, but very close to the location of the vortex the charge
approaches the bulk value and changes sign. The trend is similar to the case
studied in section II.B even though the effect is more pronounced in this last case.

\begin{figure}
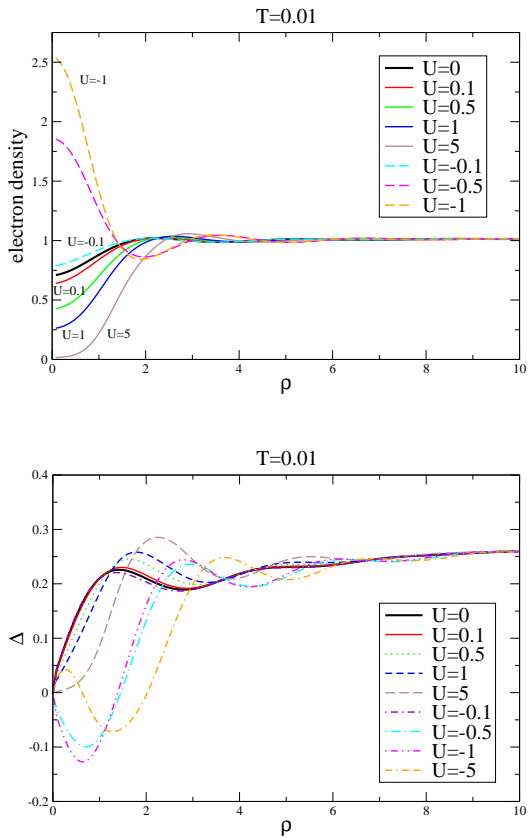

\subfigure{\includegraphics[width=0.8\columnwidth]{impurity.eps}}
\subfigure{\includegraphics[width=0.8\columnwidth]{delimp.eps}}
\caption{\label{fig20}
(color online) Electron density and $\Delta$ as a function of
distance for different values of $U$. The disc size is $d=0.96$.
}
\end{figure}

In Fig. \ref{fig20} we consider the effect of the
impurity potential at a low value of
$T$. For positive values of $U$ as $U$ increases the charge is further depleted
decreasing both the value of the electron density (it vanishes at the origin
for a sufficiently large value $U=5$) and extending the regime where the charge
is depleted. A value of $U=5$ is again similar to any larger value.
If the impurity potential is negative ($U<0$) even small values of $U$
like, for instance, $U=-0.5$ have a considerable effect on the vortex charge.
The attractive potential accumulates charge at the vortex core.
The effect
of the impurity potential on the gap function if $U>0$ is also clear.
Increasing $U$ is similar to the
effect of increasing the temperature (Kramer and Pesch effect).
However, an attractive potential has a more profound effect. Even for a small
value $U=-0.5$ a node appears in a way similar to the node of the vortex with
two flux quanta ($m=2$) previously considered. As $U$ decreases further for instance for
$U=-5$, two nodes appear in the gap function. Therefore, there is a similarity
between the attractive impurity case, the vicinity of two vortices and the multiple-flux vortex.
The quantitative
effect on the vortex charge is however different, since the impurity potential
is more effective in changing the signal of the vortex charge.

\section{Conclusions}

Earlier treatments predicted an universal charge depletion at the vortex cores.
Taking into account the competition in the d-wave case with other order parameters
it has been determined that
in some circumstances the vortices may be negatively charged (charge
accumulation with respect to the bulk value).

In this work we did not consider the effect ot other orderings but considered
the influence of disorder. We focused on the effects of positional disorder
of the vortices and on the effect of impurities. When two vortices are close by
we found that strong fluctuations appear in the shared region of the vortices,
that induce a smaller charge accumulation. Also, the addition of impurities
changes the charge profiles. A small to moderate attractive potential
also changes the signal of the vortex charge, since it renormalizes locally
the chemical potential in a straightforward way.

The case of a vortex lattice in a very high magnetic field, where the
quasiparticles propagate coherently throughout the system and gapless
superconductivity occurs, is a qualitative different state.
In this regime a Landau level description is adequate and a different behavior
is found for many physical properties \cite{RMP}. Preliminary results seem to
indicate that the effect of the coherence on the vortex charge is to change
the signal of the vortex charge: close to the vortices there is a charge
accumulation instead of a charge depletion \cite{ps}. These results would then
be in disagreement with possible explanations of the Hall anomaly as due to
the positively charged vortices. Indeed the Hall anomaly is detected close to the
normal phase, where in strongly type-II superconductors, like high-$T_c$ materials,
it is predicted that a Landau description should be appropriate \cite{PRL1}.

\begin{acknowledgments}
We thank discussions with P. Bicudo and M. Cardoso.
\end{acknowledgments}

\end{document}